\documentstyle[11pt]{article}
\newcommand{\BE}{\begin{equation}}
\newcommand{\EE}{\end{equation}}
\newcommand{\BA}{\begin{eqnarray}}
\newcommand{\EA}{\end{eqnarray}}

\begin{document}
\begin{titlepage}

\vspace*{1mm}
\begin{center}

            {\LARGE{\bf Newtonian gravity from the Higgs field: \\ 
                        the sublimation of aether }}

\vspace*{14mm}
{\Large  M. Consoli }
\vspace*{4mm}\\
{\large
Istituto Nazionale di Fisica Nucleare, Sezione di Catania \\
Corso Italia 57, 95129 Catania, Italy}
%\vspace*{3mm}\\
%and \\
%\vspace*{3mm}
%{\Large F. Siringo}
%\vspace*{2.5mm}\\
%{\large Dipartimento di Fisica dell' Universit\`a di Catania \\
%Corso Italia 57, 95129 Catania, Italy}
%\vspace{7mm}\\
\end{center}
\begin{center}
{\bf Abstract}
\end{center}

We illustrate why
a space-time structure as in General Relativity is not in contradiction with
a dynamical origin of gravity from a scalar field. Further, 
we argue that the recently discovered
gap-less mode of the singlet Higgs field represents the most natural 
dynamical agent of Newtonian gravity. 
\vskip 35 pt
%\par KEYWORDS: Gravitation, Relativity, Vacuum-condensation.
\end{titlepage}

{\bf 1.}~~Following the original induced-gravity models \cite{adler}, one may
attempt to describe gravity as
an effective force induced by the vacuum structure, in analogy
with the attractive interaction among electrons that can only
exist in the presence of an ion lattice. Looking for such a
 `dynamical' mechanism one should also be able to account for
those peculiar `geometrical' properties at the base of the classical space-time
description of General Relativity.

The idea that gravitation requires somehow
an underlying {\it scalar} field arises naturally when considering
the non-trivial ambiguity \cite{fulton}
associated with the operative definition of the 
infinitesimal transformation to the reference frame of a freely falling
observer. Indeed, within General Relativity, the
Equivalence Principle is used to relate a frame in which 
a constant force is acting to a frame in a Riemannian space 
uniformly accelerating with respect to an inertial frame. An alternative
and equivalent description \cite{fulton} is obtained by considering 
infinitesimal acceleration transformations of the conformal group. In this
case, the equivalence is not with a frame in a Riemannian space but 
with a frame in a more general Weyl space, i.e. a space
where the affine connections are not given by the Christoffel symbols
\cite{fulton} since at each point $x$, 
besides a symmetric tensor $g_{\mu \nu} (x)$, there
is a vector $\kappa_{\mu}(x)$. 
One usually considers the Weyl space in relation to
the original Weyl's theory including electromagnetism. That theory
predicts that the length of a rod depends on his history, which may seem
physically unteinable. However, the idea of a Weyl space 
cannot be rejected
{\it a priori} so that one is still faced with the interpretation of the 
infinitesimal acceleration
transformation and as such of the basic ingredient
underlying any space-time description in gravitational fields. In this sense, 
the formal foundation of General Relativity 
contains an arbitrary assumption, at least beyond the weak-field limit 
where it has been experimentally tested. 

In an alternative approach, one could have argued 
as follows. {\it A priori}, each of the two possibilities
represents an arbitrary choice. Therefore, it may be natural to solve
the ambiguity by adopting the more general 
conformal-transformation point of view but restricting to that special
class of Weyl spaces that are {\it equivalent} to Riemannian spaces. 
This equivalence can be expressed as $|g|\equiv 1$ and
\BE
\label{kappa}
{{\kappa_\mu }\over{4\pi}} =- \partial_\mu \tilde{\sigma}
\EE
where
$\tilde{\sigma}$ is a scalar field. In this case, in fact, the Weyl connections
constructed with $g_{\mu \nu}$ and $\kappa_\mu$ are the Christoffel symbols 
of the metric tensor \cite{fulton}
\BE
\label{det}
\hat{g}_{\mu\nu}= e^{-4 \pi \tilde{\sigma}} g_{\mu\nu}
\EE
To explore the possible implications of this space-time structure, 
let us consider the case of a static isotropic metric
\BE
\hat{g}_{\mu \nu}\equiv(A,-B,-B,-B)
\EE
for which $ AB^3=e^{-16 \pi \tilde{\sigma}}$.
Further, we can require the operative definition of the light velocity
to be equivalently obtained as the coordinate velocity
$({{dx^2 + dy^2 + dz^2}\over{dt^2}})^{1/2}=\sqrt{A/B} $ 
obtained from $ds^2=0$ or 
the group velocity of a light pulse, solution of
the covariant D'Alembert wave equation. In this case, 
one finds \cite{progress}
$AB={\rm const}$. Finally, with a flat space-time 
at infinity one gets $AB=1$ and the unique form
\BE
\label{unique}
    ds^2= e^{8 \pi \tilde{\sigma}}dt^2 - 
e^{- 8 \pi \tilde{\sigma}} (dx^2 + dy^2 + dz^2)
\EE
In principle
the scalar function $\tilde{\sigma}$, describing a given gravitational field,
can be determined {\it without solving any field equation} but simply 
comparing with experiments in the weak-field regime.
This would lead to the identification of $\tilde{\sigma}$
with the Newton potential $\phi_N$ (`EXP'=Experimental)
\BE
\label{identify}
                    \tilde{\sigma}_{\rm EXP}=- {{ \phi_N}\over{ 4 \pi}}
\EE
i.e. the solution of the Poisson equation in flat space
corresponding to a given mass density.
In this way Eq.(\ref{unique}) represents the space-time metric of an
asymptotically far observer at infinity
where the gravitational potential $\phi_N(\infty)=0$. 

As an independent argument and
to better appreciate the implications of Eqs.(\ref{unique}) and
(\ref{identify}), we shall now
follow the original idea proposed by Yilmaz long time ago \cite{yilmaz}. 
In this way, Einstein equations for the Yilmaz's metric
\BE
\label{yilmazold}
    (ds^2)_{\rm Yilmaz}= e^{-2\phi_N}dt^2 - e^{2\phi_N} (dx^2 + dy^2 + dz^2)
\EE
become algebraic identities \cite{yilmaz} when using Poisson equation.
In particular, outside
of the sources of $\phi_N$, namely where $\Delta\phi_N=0$, 
 one can freely interchange \cite{net} 
$G^\mu_\nu \equiv R^\mu_\nu -{{1}\over{2}}\delta^\mu_\nu R$ for the metric
Eq.(\ref{yilmazold}) with
the stress-tensor for the $\phi_N$ field 
\BE
\label{stress}
      t^{\mu}_{\nu}(\phi_N)= 
-\partial^\mu \phi_N\partial_\nu\phi_N
+{{1}\over{2}}\delta^{\mu}_{\nu}
\partial^\rho\phi_N\partial_\rho\phi_N
\EE
In this sense, the scalar field $\phi_N$ represents the true agent
of the gravitational interaction for static fields. Notice that there is
no contradiction between a {\it scalar} field, as a
dynamical agent, and a {\it tensor} field, namely
$ \hat{g}_{\mu \nu} = \hat{g}_{\mu \nu} (\phi_N)$, to account for
the relevant space-time effects \cite{absurd}. 

Although the overall picture looks similar
to General Relativity, there are some non-trivial differences. For instance, 
in standard General Relativity the free-fall 
transformation is understood as
a general coordinate transformation. Therefore, differently from
$ t^{\mu}_{\nu}(\phi_N)$, Einstein gravitational stress-tensor is only 
a {\it pseudotensor} \cite{erwin} that vanishes for some 
choice of coordinates but becomes non-zero with another set of coordinates.

Also, for weak field (and in the {\it one-body} case)
the space-time structure (\ref{yilmazold}) becomes equivalent to 
the Schwarzschild metric
\BE
\label{schwarz}
    (ds^2)_{\rm Schwarzschild} = [{{ 1- \phi_N/2 } \over{  1+ \phi_N/2 }} ]^2dt^2 -
(1+ \phi_N/2)^4(dx^2+dy^2+dz^2)
\EE
up to higher order terms.
However, Eq.(\ref{yilmazold}) applies to a gravitationally interacting N-body
system and does not contain black holes. As a consequence, one 
predicts different stability regimes for compact massive objects
as neutron stars \cite{yilmaznew}. At the same time, by exploiting the unique 
factorization properties of the exponential metric Eq.(\ref{yilmazold}), 
one can explain the huge quasar red-shifts at moderate distances
by purely gravitational (rather than cosmological) effects \cite{clapp}.

Concerning these differences, 
we emphasize that the metric structure Eq.(\ref{unique})
is of very general nature and it stands
by itself regardless of any field equations.
On the other hand, in the
one-body case, where the two metrics 
Eqs.(\ref{yilmazold}) and (\ref{schwarz}) 
can be compared, there is an interpolating
parameterization \cite{tupper} ($\beta=1- \epsilon^2$) 
\BE
\label{tupper}
    ds^2(\beta) = 
[{{ 1-\epsilon \phi_N/2 } \over{1+\epsilon \phi_N/2 }} ]
^{2/\epsilon} dt^2 -
[{{ 1+\epsilon \phi_N/2 } \over{1-\epsilon \phi_N/2 }} ]
^{2/\epsilon} (1- {{\epsilon^2 \phi^2_N}\over{4}} )^2(dx^2+dy^2+dz^2)
\EE
that reduces to Eq.(\ref{yilmazold}) for $\beta=1$ and 
to Eq.(\ref{schwarz}) for $\beta=0$. In this case, 
a sufficiently precise experiment should objectively
determine the numerical value of $\beta$. It is surprising, though, 
that according to refs.\cite{progress,annals} one could already use
existing measurements, namely the
experiments on neutron phase shift in a gravitational field \cite{colella} 
and those on the isotropy of inertia \cite{drever}, to deduce 
$|1-\beta|= {\cal O} (10^{-2})$ thus ruling out the 
Schwarzschild metric Eq.(\ref{schwarz}).

Beyond the static case, Yilmaz's original approach can be considered
the limit of a framework 
where the space-time structure is distorted by 
several fields. Now
Eqs.(\ref{kappa}) and (\ref{det}) hold for the general forms
\BE
\label{wealth}
\hat{g}_{\mu\nu}=\hat{g}_{\mu\nu}
( \phi_N,A_\mu,B_{\mu\nu},..)
\EE
\BE
\label{kappa2}
     \kappa_\mu=\kappa_\mu(\phi_N,A_\mu,B_{\mu\nu},..)
\EE
where $A_\mu$, $B_{\mu\nu}$... represent unknown dynamical
agents providing possible new contributions 
(as $A_\mu A^\mu$, $B^\mu_\mu$, $B^{\mu\nu}B_{\mu\nu}$...) 
to the effective scalar $\tilde{\sigma}$. 
Although determining their precise nature would require to control
all possible aspects of gravity, one can explore some general consequence
of identifying $\phi_N, A_\mu, B_{\mu\nu}$,..as long-wavelength
excitations 
(i.e. {\it disturbances}) of the quantum field theoretical vacuum. 

This point of view arises naturally when starting from
Minkowski space to describe the
elementary particle interactions but taking into account that
the physical vacuum of quantum field theory is not trivially
`empty'. Rather
the phenomenon of vacuum condensation \cite{salehi}
should be considered the operative construction of a `quantum aether'
\cite{dirac}, i.e. different from the aether of
classical physics but also different from the `empty' space-time of special
relativity.
In this approach, the parametric dependence in Eqs.(\ref{wealth}) and
(\ref{kappa2}) is such that any deviation from a flat space-time is due to 
peculiar field configurations $\phi_N, A_\mu, B_{\mu\nu}$,.. 
In analogy with the static case and outside of the
sources of $\phi_N, A_\mu, B_{\mu\nu}$,.. the metric structure is such to
ensure the formal identity between the $G^\mu_\nu$ corresponding
to Eq.(\ref{wealth}) and a suitable `gravitational' 
stress-tensor $t^{\mu}_{\nu}(\phi_N,A,B,..)$ \cite{american}. 
Although one can get
agreement with General Relativity to first-order in
$\phi_N, A_\mu, B_{\mu\nu}$,..
($t^{\mu}_{\nu}(\phi_N,A,B,..)$ is, at least, a second-order quantity) 
the problem of the vacuum energy is very different from the standard approach
and one can easily understand the absence of any curvature
in the unperturbed ground state where, by definition, 
$t^{\mu}_{\nu}(\phi_N,A,B,..)=0$.

However, before attempting to describe
the excitation states of the vacuum in the 
most general frameworks (fastly rotating objects, sources of
gravitational waves,...)
it is natural to proceed step by step starting from the simplest case. 
For this reason, in the rest of this Letter, we shall only try to
understand the physical meaning of the scalar field
$\tilde{\sigma}$ for the well known case of
Newtonian gravity. After all, independently
of any field equations, the general space-time structure
expressed in Eqs.(\ref{kappa})-(\ref{identify}) is consistent with all 
experiments and requires a single scalar function: the Newton potential.
Looking for its dynamical origin, we shall propose the possible 
interpretation of $\tilde{\sigma}_{\rm EXP}=-{{\phi_N}\over{4\pi}}$ as a
collective excitation of the same scalar condensate at the base of mass
generation in the Standard Model of electroweak interactions. 
In the version Eqs.(\ref{kappa})-(\ref{identify})
of the Equivalence Principle, the density 
fluctuations of the physical
vacuum \cite{volo} can be described geometrically as
distortions of the flat Minkowski space-time into
a particular class of Weyl spaces. 

\vskip 7 pt

{\bf 2.}~~Our proposal is motivated by recent
theoretical arguments \cite{singlet,legendre} that, 
quite independently of the Goldstone phenomenon, suggest the existence
of a gap-less mode of the singlet  Higgs field
in the broken phase of a $\lambda \Phi^4$ 
theory. Its presence is a direct consequence of the quantum 
nature of the scalar condensate that cannot be treated as a purely 
classical c-number field. In fact, either considering the 
re-summation of the one-particle reducible
zero-momentum tadpole graphs \cite{singlet} in a given background field 
or performing explicitely the last functional integration over the strength
of the zero-momentum mode of the singlet Higgs field
\cite{legendre}, one finds {\it two} possible solutions
for the inverse zero-4-momentum propagator in the spontaneously broken phase:
a)~$G^{-1}_a(0)=M^2_h$ and b)~$G^{-1}_b(0)=0$. 

This result is not totally unexpected and admits a simple
geometrical interpretation in terms of the shape of
$V_{\rm LT}(\varphi)$, 
the effective potential obtained from the Legendre transform (`LT') of
the generating functional for connected Green's function. 
Differently from the conventional non-convex (`NC')
effective potential $V_{\rm NC}(\varphi)$, as computed in the loop
expansion, $V_{\rm LT}(\varphi)$ is not an infinitely 
differentiable function in the presence of spontaneous symmetry breaking 
\cite{syma}. Moreover, being rigorously convex downward, 
$V_{\rm LT}(\varphi)$ will agree with
$V_{\rm NC}(\varphi)$ everywhere {\it except}
in the region enclosed by the 
absolute minima of $V_{\rm NC}(\varphi)$, say $\varphi=\pm v$, where it
is exactly flat. For this reason, although
an exterior second derivative of
$V_{\rm LT}(\varphi)$ will agree with
$V''_{\rm NC}(\pm v)\equiv M^2_h$, one will also be faced with a vanishing
internal second derivative which has no counterpart in 
$V_{\rm NC}(\varphi)$.

As anticipated, the main point of Refs.\cite{singlet,legendre}
is that the `Maxwell construction', 
i.e. the replacement $V_{\rm NC}(\varphi) \to 
V_{\rm LT}(\varphi)$ reflects the
quantum nature of the scalar condensate
(see Ref.\cite{branchina} for a still further derivation). As such, 
there are important consequences for the infrared behaviour of the
theory.
In fact, two possible values for the zero-4-momentum propagator imply
two possible types of excitations with different energies when the 3-momentum
${\bf{p}} \to 0$: 
a massive one, with $\tilde{E}_a({\bf{p}}) \to M_h$, and a gap-less one with 
$\tilde{E}_b({\bf{p}}) \to 0$. The latter would
clearly dominate the far infrared region ${\bf{p}} \to 0$ where the massive
branch becomes unphysical. 
Therefore, differently from the most naive perturbative indications, 
in a (one-component) spontaneously broken
$\lambda\Phi^4$ theory there is no energy-gap associated
with the `mass'  $M_h$
of the shifted field $h(x)=\Phi(x) -\langle\Phi\rangle$, 
as it would be for a genuine massive single-particle 
spectrum where the relation
\BE
\label{Ea}
\tilde{E}({\bf{p}})=\sqrt{ {\bf{p}}^2 + M^2_h} 
\EE
remains true for ${\bf{p}} \to 0$. Rather, the infrared region
is dominated by gap-less collective modes
\BE
\label{Eb}
\tilde{E}({\bf{p}}) \equiv c_s |{\bf{p}}| 
\EE
depending on an unknown parameter $c_s$ which controls the slope of the spectrum
for  ${\bf{p}} \to 0$ and has the meaning of the `sound velocity' 
for the density fluctuations of the scalar condensate.
Indications on its magnitude can be obtained on the base of a semi-classical
argument due to Stevenson \cite{seminar} that we shall briefly report below. 

Stevenson's argument starts from a perfect-fluid treatment of the Higgs
condensate. In this approximation, 
energy-momentum conservation is equivalent to wave propagation with a
squared velocity given by ( $c$ is the light velocity)
\BE
\label{cs}
        c^2_s= c^2 ({{\partial P}\over{\partial {\cal E} }})
\EE
where $P$ is the pressure and ${\cal E}$ the
energy density. Introducing the condensate density $n$, and using
the energy-pressure relation
\BE
\label{zerot}
 P= -{\cal E} + n {{\partial {\cal E} }\over{\partial n}}
\EE
we obtain
\BE
\label{nn1}
c^2_s=
c^2({{\partial P}\over{\partial n}})
({{\partial {\cal E} }\over{\partial n}})^{-1}=
c^2  (n{{ \partial^2 {\cal E} }\over{\partial n^2}})
({{\partial {\cal E} }\over{\partial n}})^{-1} 
\EE
For a non-relativistic Bose condensate of neutral particles with mass
$m$ and scattering length $a$, where ${{na\hbar^2}\over{m^2c^2}} \ll 1$
(in this case we explicitely introduce $\hbar$ and $c$) one finds
\BE
\label{nonrel}
{\cal E}= n m c^2 + n^2 {{2\pi a \hbar^2}\over{m}} 
\EE
so that 
\BE
\label{nn2}
c^2_s={{4\pi n a \hbar^2}\over{m^2}}
\EE
which is the well known result for the sound velocity in a  
dilute hard sphere Bose gas \cite{huang}.

 On the other hand, in a fully relativistic case, the additional terms
in Eq.(\ref{nonrel}) are such that the scalar condensate 
is spontaneously generated from the `empty' vacuum where $n=0$ for
that particular equilibrium density where \cite{mech}
\BE
\label{vacuum}
{{\partial {\cal E} }\over{\partial n}} = 0
\EE
Therefore, in this approximation, approaching the equilibrium density
one finds
\BE
\label{infinity}
           c^2_s \to \infty
\EE
thus implying that long-wavelength
density fluctuations would propagate instantaneously in the 
spontaneously broken vacuum. 

As Stevenson points out \cite{seminar}, 
Eq.(\ref{infinity}) neglects all possible corrections to the perfect-fluid
approximation, as well as finite-temperature effects (as for instance if
the Higgs condensate were in thermal equilibrium with the microwave 
background radiation). Still, the above semi-classical argument suggests       
that density fluctuations can propagate
extremely fast in the spontaneously broken vacuum, 
at least in the long-wavelength limit
where the identification (\ref{Eb})
$c_s={{d\tilde{E}}\over{ d|{\bf{p}}| }}$  applies. 

On the other hand, 
sound waves cannot propagate with too short 
wavelengths. In our case there is a typical momentum, say
$|{\bf{p}}|= \delta$, of the order of the inverse mean 
free path for the elementary costituents in the scalar
condensate, where the collective
modes become unphysical and a single-particle
spectrum as in Eq.(\ref{Ea}) applies. The transition occurs where
\BE
\label{beyond}
\sqrt{ \delta^2 + M^2_h} \sim c_s \delta
\EE
so that, for $c_s \to \infty$, $\delta$ is naturally infinitesimal in units
of $M_h$. For this reason 
superluminal wave propagation is restricted to the region 
${\bf{p}} \to 0$ and does not necessarily implies 
violations of causality as for a group velocity
${{d\tilde{E}}\over{ d|{\bf{p}}| }} >1 $ when $|{\bf{p}}| \to \infty$ 
\cite{seminar}.
\vskip 7 pt
{\bf 3.}~~Independent informations on $c_s$ can be obtained by comparing with 
phenomenology.
Let us ignore, for the moment, the previous indication in Eq.(\ref{infinity})
and just explore the consequences of Eq.(\ref{Eb}).
To this end, let us define $\tilde{h}(x)$ to be the component
of the fluctuation field associated with the long-wavelength modes
Eq.(\ref{Eb}). Whatever the value of $c_s$, 
these dominate the infrared region so that
a general yukawa coupling of the Higgs field to fermions will give
rise to a long-range {\it attractive} potential
between any pair of fermion masses $m_i$ amd $m_j$ 
\BE
\label{Newton}
            U_{\infty}(r)= - {{1}\over{4\pi c^2_s
\langle \Phi \rangle^2 }}{{m_im_j}\over{r}}
\EE
The above result would have a considerable impact
for the Standard Model if we take the value $\langle\Phi\rangle \sim 246$ GeV
related to the Fermi constant. Unless $c_s$ is an extremely 
large number (in units of $c$) one is faced with strong
long-range forces coupled to the inertial masses of the known elementary 
fermions. Just to have an idea, by assuming $c_s = 1$ 
the long-range interaction between two electrons in
Eq.(\ref{Newton}) would be ${\cal O}(10^{33})$ larger than their purely
gravitational attraction. 
On the other hand, invoking a phenomenologically viable strength,  
as if $c_s \langle\Phi\rangle$ would be
of the order of the Planck scale, is equivalent to re-obtain 
nearly instantaneous interactions transmitted by the scalar condensate as 
in Eq.(\ref{infinity}).

On a macroscopic scale, the relevant long-range effects can be evaluated 
from the effective lagrangian 
${\cal L}_{\rm eff}$ for  $\tilde{h}(x)$. 
In an expansion in powers of
$\tilde{h}(x)$, the lowest-order interaction term
arises from its coupling to the trace 
of the energy-momentum tensor of ordinary matter
$T^{\mu}_{\mu}(x)$ and represents the obvious
renormalization of the elementary yukawa couplings 
\BE
        \langle f| T^{\mu}_{\mu}| f \rangle=
         m_f \bar{\psi}_f \psi_f
\EE
In this way, by defining
$\tilde{\sigma}\equiv {{\tilde{h}}\over{\langle\Phi\rangle}} $ we get
\BE
\label{lagrangian}
 {\cal L}_{\rm eff} (\tilde{\sigma})=  
{{ \langle\Phi\rangle^2}\over{2}} \tilde{\sigma}
       [ c^2_s \Delta - {{\partial^2 } \over{\partial t^2}}] \tilde{\sigma}
      - T^{\mu}_{\mu} \tilde{\sigma} +...
\EE
and the linearized equation of motion
\BE
\label{step1}
       [ c^2_s \Delta - {{\partial^2 } \over{\partial t^2}}] \tilde{\sigma}=
         { { T^{\mu}_{\mu} }\over{ \langle \Phi \rangle^2}}
\EE
Now, by assuming very large values of $c_s$ 
the $\tilde{\sigma}$ Green's function has practically no retardation effects
and Eq.(\ref{step1}) describes an instantaneous interaction
\BE
\label{step2}
          \Delta \tilde{\sigma}= 
{{T^{\mu}_{\mu}}\over{c^2_s \langle\Phi\rangle^2}} 
\EE
Finally for slow motions, when the trace of the energy-momentum tensor 
\BE
         T^{\mu}_{\mu}(x) \equiv \sum_n 
{ { E^2_n -   {\bf{p}}_n  \cdot  {\bf{p}}_n  } \over{E_n}} 
\delta^3 ( {\bf{x}} - {\bf{x}}_n(t) ) 
\EE
reduces to the mass density
\BE
\label{source}
         \rho(x ) \equiv \sum_n m_n
\delta^3( {\bf{x}} - {\bf{x}}_n(t)) 
\EE
one gets
\BE
\label{poisson}
            \Delta \tilde{\sigma}= 
{{1}\over{c^2_s \langle\Phi\rangle^2}}~ \rho( x)
\EE
Therefore, if
the only free parameter of our analysis $c_s$ would be 
fixed by the relation
\BE
\label{GN}
            c^2_s= 
{{1}\over{G_N \langle\Phi\rangle^2}}= 
{\cal O}(10^{33})
\EE
$G_N$ being the Newton constant, 
Eq.(\ref{poisson}) would
become formally identical to the Poisson equation for the Newton potential
allowing for the identification in Eq.(\ref{identify}). 
As anticipated, Stevenson's Eq.(\ref{infinity}) provides a clue to understand
 the very large number
in Eq.(\ref{GN}) leading to
$c_s \sim 4\cdot 10^{16} $. Although this may seem an infinitely large value, 
$c_s$ is not so fantastically higher than the experimental lower 
limit $2\cdot 10^{10}$ 
reported by Van Flandern \cite{tom} for the `speed of gravity' when gravity is
described as a {\it central} interaction \cite{carlip} as in Eq.(\ref{poisson}).

Before concluding, we comment on 
the momentum $\delta$ Eq.(\ref{beyond}) associated with the 
transition between the two branches of the
spectrum Eqs.(\ref{Eb}) and (\ref{Ea}) for which Eq.(\ref{GN}) implies
\BE
\label{match}
 {{\delta}\over{M_h}} = {\cal O} (10^{-16})
\EE
At distances $r \sim \delta^{-1}$ the 
interparticle potential is not given by Eq.(\ref{Newton}) but
has to be computed from the 
Fourier transform of the $h-$field propagator
\BE
D(r)=
\int {{d^3 {\bf{p}} }\over{(2\pi)^3 }}
 {{e^{ i {\bf{p}}\cdot {\bf{r}} } }\over{ 
\tilde{E}^2({\bf{p}}) }}
\EE
and depends on the detailed form of the spectrum that interpolates between
Eqs.(\ref{Eb}) and (\ref{Ea}). 
However in the Standard Model, for $M_h={\cal O}(\langle\Phi\rangle)$, 
one predicts in any case a length scale $R=\delta^{-1}$ in the millimeter
range.
This length scale would become infinitely large in the limit where the 
electroweak scale
$M_{\rm EW}\equiv \sqrt{M_h\langle\Phi\rangle}$ is kept fixed and one takes
the limit Eq.(\ref{infinity}) where
$M_{\rm Planck}\equiv c_s \langle \Phi \rangle \to \infty $.
This last phenomenological feature is also found
in approaches involving extra `large' space-time dimensions \cite{dimo}
although we have no obvious explanation for this analogy.

\vskip 7 pt
{\bf 4.}~~Summarizing: the 
Equivalence Principle does not fix uniquely the structure of space-time. 
In the alternative picture
Eqs.(\ref{kappa})-(\ref{unique}), it becomes natural to look for the dynamical 
interpretation of the scalar field $\tilde{\sigma}(x)$ associated with
Newtonian gravity.  Our proposal is
to identify $\tilde{\sigma}_{\rm EXP}=-{{\phi_N}\over{4\pi}}$ 
as a long-wavelength excitation
of the same scalar condensate that induces spontaneous symmetry 
breaking for electroweak interactions. In this way one gets an alternative, 
but consistent, space-time description
in weak gravitational fields where the differences between
Eqs.(\ref{yilmazold}) and (\ref{schwarz}) should presently be
unobservable. We emphasize, however, that our picture
would uniquely be singled out 
by accepting the arguments of refs.\cite{progress,annals} for a
clear experimental evidence in favour of the Yilmaz metric
Eq.(\ref{yilmazold}). 

In addition, our proposal represents a natural and
`economical' way to account for
the gap-less mode of the singlet Higgs field Eq.(\ref{Eb}). Its existence
reflects the nature of the scalar condensate as a truly
physical medium responsible for the deviations from exact
Lorentz covariance in the long-wavelength 
part of the energy spectrum. For this reason, 
introducing alternative phenomenological frameworks, 
if possible, would certainly not represent the simplest logical
possibility. Moreover, by taking seriously
the basic idea that gravity is a long-wavelength excitation
of the quantum field theoretical vacuum
it will be possible to re-consider long-standing problems (quasar red-shifts, 
cosmological constant, missing mass, fifth force,...)
from a new and unifying perspective. For instance, 
a hydrodynamical solution \cite{beke} of the mass discrepancy in galactic 
systems is equivalent to introduce deviations from that particular
constant-density Poisson flow represented by Newtonian gravity. 
Such modifications are very natural \cite{consoli} in an approach where 
gravity is associated with the density fluctuations of a physical medium.
\vskip 7 pt
{\bf Acknowledgements}~~~I thank P. M. Stevenson for 
useful discussions and collaboration. I also thank A. Garuccio for informations
on the Yilmaz picture of gravity.

\end{document}